\documentclass[a4paper,12pt]{article}

\usepackage{amsmath,amssymb}

\newtheorem{proposition}{Proposition}
\newtheorem{corollary}{Corollary}
\newtheorem{theorem}{Theorem}
\newtheorem{remark}{Remark}

\title{Consumption and Portfolio Rules for Time-Inconsistent Investors
\footnote{Acknowledgements:
This work has been partially supported
by MEC (Spain) Grant MTM2006-13468}}
\author{Jes\'{u}s Mar\'{i}n-Solano\footnote{Corresponding author:
Jes\'{u}s Mar\'{i}n-Solano, Dept. Matem\`{a}tica econ\`{o}mica, financera
i actuarial, Universitat
de Barcelona, Avda. Diagonal 690, E-08034 Barcelona, Spain.
E-mail address: jmarin@ub.edu; Tel.: +34-93-402-1991; fax: +34-93-403-4892}
\, and Jorge Navas\\
{\small Dept.
Matem\`{a}tica econ\`{o}mica, financera i actuarial, Universitat de Barcelona}
\\
{\small Av. Diagonal 690, E-08034 Barcelona, Spain}\\}

\date{}

\begin{document}
\maketitle

\begin{center}
First version: October 25, 2007 \\
This version: March 4, 2009
\end{center}

\begin{abstract}
This paper extends the classical consumption and portfolio rules model
in continuous time (Merton 1969, 1971) to the framework of decision-makers
with time-inconsistent preferences. The model is sol\-ved for different
utility functions for both, naive and sophisticated agents, and the results
are compared. In order
to solve the problem for sophisticated agents, we derive a modified
HJB (Hamilton-Jacobi-Bellman) equation. It is illustrated how for CRRA functions within the family of
HARA functions (logarithmic and
potential cases) the optimal portfolio rule does not depend on the discount
rate, but this is not the case for a general utility function, such as
the exponential (CARA) utility function.
\end{abstract}


\section{Introduction}

Variable rate of time preferences have received considerable attention
in recent years. Virtually every experimental study on time preferences
suggests that the standard assumption of time-consistency (related to
the assumption of constant discount rate of time preference) is
unrealistic (see, for instance, Thaler (1981), Ainslie  (1992) or
Loewenstein and Prelec (1992)). In fact, there is substantial
evidence that agents are impatient about choices in the short term but
are patient when choosing between long-term alternatives.

Effects of the so called hyperbolic discount functions, introduced by
Phelps and Pollak (1968), have been extensively studied in a discrete
time context, within the field of behavioral economics (for a recent application
in the economics of information systems, see Tomak and Keskin (2008)). Laibson (1997)
has made compelling observations about ways in which rates of time
preference vary. However, this topic has received less attention in a
continuous time setting. The main reason for this may be the complexity
involved in the search for solutions in closed form in the non-constant
discounting case. In fact, standard optimal control techniques cannot be
used in this context, since they give rise to non-consistent policies.

The most relevant effect of non-constant discounting is that preferences
change with time. An agent making a decision in
time $t$ has different preferences compared with those at time $t^\prime$.
Therefore, we can consider him or her at different times as different agents.
An agent making a decision at time $t$ is usually called the
$t$-agent. If the planning horizon is a finite interval $[0,T]$, we can
understand the dynamic optimization problem with non-constant discounting as
a perfect information sequential game with a continuous number of players
(the $t$-agents, for $t\in [0,T]$) making their decisions sequentially.
A $t$-agent can act in two different ways: naive and sophisticated.

Naive agents take decisions without taking into account that their
preferences will change in the near future. Then, they will be
continuously modifying their calculated choices for the future, and their
decisions will be in general time-inconsistent. In order to obtain a time
consistent strategy, the $t$-agent should be
sophisticated, in the sense of taking into account the preferences of all
the $t^\prime$-agents, for $t^\prime\in(t,T]$. Therefore, the solution to
the problem of the agent with non-constant discounting should be constructed
by looking for the subgame perfect equilibria of the associated game with
an infinite number of $t$-agents.

Historically, in that part of his analysis allowing for time
preference, Ramsey (1928) assumed an exponential discount factor with constant
discount rate, stating: ``This is the only assumption we can make, without
contradicting our fundamental hypothesis that successive generations are
activated by the same system of preferences''. The main property of non-constant
discounting is implicit in this statement: it can create a time-consistency
problem. In fact, Strotz (1956) illustrated how, for a very simple model,
preferences are time consistent if, and only if, the discount factor
representing time preferences is an exponential with a constant discount rate.
In order to avoid such time inconsistency,
agents could decide in a sophisticated way, making an analysis of what
their actions would be in the future, as a consequence of their changing preferences.
For instance, Pollak (1968) gave the right solution to the Strotz problem for both
naive and sophisticated agents under a logarithmic utility function.

Although the problem was first presented in a continuous time context (Strotz
(1956)), almost all attention has been given to the discrete time
setting introduced by Phelps and Pollak (1968). This is probably a consequence
of the non-existence of a well-stated system of equations giving
a general method for solving the problem, at least for sophisticated agents.
Therefore, each particular problem has been solved individually.
This was the case of the Strotz model solved by Pollak in 1968.
Barro (1999) studied a modified version of the
neoclassical growth model by including a variable rate of time preference.

For the case of naive agents, one should solve a standard optimal control
problem for each time $t\in[0,T]$, in order to find the decision rule at
time $t$ of a $t$-agent. Unfortunately, this method cannot be used if the
agent is sophisticated. Instead, Markov subgame perfect equilibria
must be found. This prompts the use of a dynamic programming approach,
applying the Bellman optimality principle.

To solve the intra-personal game for sophisticated agents, a
continuous-time model of quasi-hyperbolic time preferences was introduced
in Harris and Laibson (2008). In their model, the discount rate declines
during the first ``period'' (instantaneous gratification), and then becomes
constant. Grenadier and Wang (2007) employed this model to extend the real
options framework in order to analyze the investment-timing decisions (in
an irreversible investment framework) of entrepreneurs with time-inconsistent
preferences.

In a deterministic environment, Karp (2007) adapted the approach by Harris and
Laibson (2008) for the general case of an arbitrary discount rate of time
preference, for autonomous infinite time horizon problems. The free terminal
time case in non-autonomous problems in finite horizon was analyzed in
Mar\'{\i}n-Solano and Navas (2009). An alternative
approach to the problem was given in Ekeland and Lazrak (2008).

In this paper we extend the results by Karp (2007) and Mar\'{\i}n-Solano and
Navas (2009) to a stochastic environment, in order to analyze how
time-incon\-sis\-tent
preferences modify the classical optimal consumption and portfolio rules when
the discount rate is constant (Merton (1969), (1971)). As
expected, the rate of time preference plays no role in a pure optimal portfolio
management problem. However, if the consumption is introduced in the model, an
inter-temporal conflict arises. We show that, within the HARA (hyperbolic
absolute risk aversion) functions, if the relative risk aversion is constant
(logarithmic and potential utility functions), the optimal portfolio rule
does not depend on the rate of time preference, although the consumption
rule changes. This nice property is not satisfied for more general
utility functions, such as the (constant absolute risk aversion) exponential
function.

The paper is organized as follows. In Section 2 we describe the model.
The general stochastic dynamic optimization
problem with non-constant discount rate of time preference is studied in
Section 3, and the dynamic programming (Hamilton-Jacobi-Bellman) equation
is derived. In Section 4, this equation is solved for the ``optimal''
(in fact, equilibrium) consumption and portfolio rules problem for some
particular utility functions. The so-called pre-commitment, naive and
sophisticated solutions are compared. Finally, Section 5 contains the main conclusions of the paper.

\section{The Model}

Let $x=(x^1,\dots,x^n)\in X$ be the vector of state variables, $u=(u^1,\dots,u^m)\in U$
the vector of control (or decision) variables, ${\cal L}(x(s),u(s),s)$
the instantaneous utility function at time $s$, and $F(x(T))$ the final (bequest) function.
In the conventional model, agent
preferences at time $t$ take the form
\begin{equation}\label{standard}
U_t = E\left[ \int_t^T e^{-\rho(s-t)} {\cal L}(x(s),u(s),s)\, ds +
e^{-\rho(T-t)}F(x(T))\right]\; ,
\end{equation}
where the state variables evolve according to the
diffusion equations
\begin{equation}\label{DE}
d x^i(s)=f^i(x(s),u(s),s) ds + \sum_{l=1}^L \bar{\sigma}^i_l(x(s),u(s),s)dw^l(s)\; ,
~x^i(t)=x^i_t\; ,
\end{equation}
for $i=1,\dots,n$, where $(w^1(s),\dots,w^L(s))$ is an $L$-dimensional
Wiener process with independent components ($d w^l(s) d w^{l^\prime}(s)=0$,
for $l\neq l^\prime$). We will denote $\bar{\Sigma}=\left(\bar{\sigma}^i_l
\right)$, for $i=1,\dots n$, $l=1,\dots,L$. In order to maximize $U_t$, we must
solve a stochastic optimal control problem, and since the discount rate
is constant, the solution becomes time consistent.

Now, following Karp (2007), let us assume that the instantaneous discount rate
is non-constant, but a function of time $r(s)$, for $s\in[0,T]$. Impatient
agents will be characterized by a non-increasing discount
rate $r(s)$. The discount factor at time $t$ used to evaluate
a payoff at time $t+\tau$, $\tau\geq 0$, is $\theta(\tau)= \exp\left( -\int_0^\tau r(s)\,
ds\right)$. Then, the objective of the agent
at time $t$ (the $t$-agent) will be
\begin{equation}\label{ND}
\max_{\{ u(s)\}} E \left[
\int_t^T \theta(s-t) {\cal L}(x(s),u(s),s)\, ds + \theta(T-t)F(x(T)) \right] \; .
\end{equation}

In Problem (\ref{DE}-\ref{ND}), we assume the usual regularity conditions,
i.e., functions ${\cal L}$, $F$, $f^i$ and $\sigma^i_j$ are continuously
differentiable in all their arguments.

In the discrete time case, most papers work with the
so-called hyperbolic discounting, first proposed by Phelps and Pollak
(1968). The utility function is defined as $U_t = u_t +
\beta(\delta u_{t+1} + \delta^2 u_{t+2} + \delta^3 u_{t+3}+\cdots)$,
where $0<\beta\leq 1$, and $u_k$ denotes the utility in period $k$.
In fact, Laibson (1997) argues that $\beta$
would be substantially less than one on an annual basis, perhaps
between one-half and two-thirds. Harris and Laibson (2008) adapted
this inter-temporal utility function to the continuous time setting.
As a natural extension of the above discount
function to the continuous setting, Barro (1999) suggested the
instantaneous discount rate $r(\tau)=\rho+b e^{-\gamma\tau}$,
where $b\geq 0$ and $\gamma>0$ (for the general case, he defined
the discount factor as $\theta(\tau)=e^{-[\rho\tau+\phi(\tau)]}$).
In other applications it is natural to
assume that the discount factor is a linear combination of exponentials with
constant but different discount rates. We will not assume
any particular discount function. 

In this paper we are interested in the optimal consumption and
portfolio rules in continuous time studied by Merton (1969, 1971). Let us
assume that there are $m$ risky assets and one risk-free asset.
The risk-free asset pays a
constant rate $\mu_0$, while the return of $i$-th risky asset follows a
geometric Brownian motion
\[
d P_i=\mu_i P_i ds + \sigma_i P_i dz_i\; ,~~i=1,\dots,m\; ,
\]
with $dz_i dz_j=\rho_{ij}dt$ for $i,j=1,\dots m$, $i\neq j$, and
$\mu_i,\sigma_i$ are constants.

If $w_i$ is the share of wealth invested in the $i$-th risky asset,
and $c$ denotes the consumption, the consumer's budget equation is
\begin{equation}\label{budget}
d W = \left[\sum_{i=1}^{m} w_i(\mu_i-\mu_0)W+(\mu_0 W-c)\right] ds
+ \sum_{i=1}^{m} w_i\sigma_i W dz_j\; ,
\end{equation}
with the initial condition $W_0$. Then the consumer-investor's problem is
\begin{equation}\label{CI}
\max_{\{ c,w_j\}} E \left[
\int_t^T \theta(s-t) u(c_s)\, ds + \theta(T-t)F(x(T)) \right]
\end{equation}
s.t. (\ref{budget}) with the initial condition $W(t)=W_t$.

\section{Dynamic Programming Equation}

For the solution of Problem (\ref{DE}-\ref{ND}) (and, in particular,
Problem (\ref{budget}-\ref{CI})), if the agent is naive, then we can adapt
the standard techniques of stochastic optimal control theory as
follows. If $V^0=V^0(x,s)$ is the value function,
the $0$-agent will solve the standard Hamilton-Jacobi-Bellman
equation
\[
r(s)V^0-V^0_s = \max_{\{ u\}}
\left\{ {\cal L}+V^0_x\cdot f + \frac{1}{2}tr\left(\bar{\Sigma}\bar{\Sigma}^\prime V^0_{xx}\right)
\right\}\; ,~x(0)=x_0\; ,
\]
i.e., the naive agent at time $0$ solves the problem, assuming that the
discount rate of time preference will be $r(s)$, for $s\in[0,T]$. In the
equation above we denote $V^0_x=\left(
\frac{\partial V^0}{\partial x^1},\dots,\frac{\partial V^0}{\partial x^n}\right)$,
and $V^0_{xx}=\left(\frac{\partial^2 V^0}{\partial x^i
\partial x^j}\right)$, for $i,j=1,\dots n$. The
optimal control will be a function $u^0(s)$. In our framework of
changing preferences, this solution corresponds to the so-called
pre-commitment solution, in the sense that it is optimal as long
as the agent can precommit (by signing a contract, for example)
his or her future behavior at time $t=0$, and it will be denoted by
$V^0=V^P$. If there is no commitment,
the $0$-agent will take the action $u^0(0)$ but, in the near future,
the $\epsilon$-agent will change his decision rule (time-inconsistency)
to the solution of
\[
r(s-\epsilon)V^{\epsilon}-V^{\epsilon}_s = \max_{\{ u\}}
\left\{ {\cal L}+V^{\epsilon}_x\cdot f + \frac{1}{2}tr\left(\bar{\Sigma}{\bar\Sigma}^\prime
V^{\epsilon}_{xx}\right)
\right\}\; ,~x(\epsilon)=x_\epsilon\; .
\]
Once again the optimal control trajectory $u^\epsilon(s)$, $s\in[\epsilon,T]$
will be changed for $s>\epsilon$ by the following $s$-selves. In general,
the solution for the naive agent will be constructed by solving the
family of HJB equations
\[
r(s-t)V^t-V^t_s = \max_{\{ u\}}
\left\{ {\cal L}+V^t_x\cdot f + \frac{1}{2}tr\left(\bar{\Sigma}\bar{\Sigma}^\prime V^t_{xx}\right)
\right\}\; ,~x(t)=x_t\; ,
\]
for $t\in[0,T]$, and patching together the ``optimal'' solutions $u^t(t)$.

If the agent is sophisticated, things become more complicated. The
standard HJB equation cannot be used to construct the solution,
and a new method is required. In what follows, we will derive
a modified HJB equation which will help us to find the solution
to Problem (\ref{DE}-\ref{ND}) and then (\ref{budget}-\ref{CI}).

\subsection{The discrete time case}

We will derive first the dynamic programming equation for
a discretized version of Problem (\ref{DE}-\ref{ND}), following a
procedure similar to the one used in Karp (2007) and Mar\'{\i}n-Solano
and Navas (2009).

Let us divide the interval $[0,T]$ into $N$ periods of
constant length $\epsilon$, in such a way that we identify
$d s = \epsilon$, and $s=j\epsilon$, for $j=0,1,\dots,N$.
Denoting by $x(j\epsilon)=x_j$, $u(k\epsilon)=u_k$ ($j,k=0,\dots,N-1$),
the objective of the agent in period $t=j\epsilon$ will be
\begin{equation}\label{Dmax}
\max_{\{ u_k\}} V_j = E\left[\sum_{i=0}^{N-j-1} \theta(i\epsilon) {\cal L}(x_{i+j},
u_{i+j},(i+j)\epsilon)\epsilon + \theta((N-j)\epsilon)F(x(T))\right]\; ,
\end{equation}
\begin{equation}\label{DCE}
x^i_{k+1} = x^i_k + f^i(x_k,u_k,k\epsilon)\epsilon +
\sum_{l=1}^L\bar{\sigma}^i_l(x_k,u_k,k\epsilon)( w^l_{k+1} - w^l_k
) \; ,
\end{equation}
for $i=1,\dots n$ and $k=j,\dots,N-1$, with $x_j$ given.

Let us state the dynamic programming algorithm for the discrete
problem (\ref{Dmax}-\ref{DCE}). In the final period, $t=N\epsilon=T$,
we define $V_N^*=F(x_N)$. For $j=N-1$, the optimal
value for (\ref{Dmax}) will be given by the solution to the problem
\[
V_{(N-1)}^* =  \max_{\{u_{N-1}\}}
E\left[ {\cal L}(x_{N-1}, u_{N-1},(N-1)\epsilon)\epsilon + \theta_{1} V_N^*\right]\; ,
\]
with $x^i_{N} = x^i_{N-1} + f^i(x_{N-1},u_{N-1},(N-1)\epsilon)\epsilon +
\sum_{l=1}^L\bar{\sigma}^i_l(x_{N-1},u_{N-1},(N-1)\epsilon)( w^l_{N} -
w^l_{N-1})$, for $i=1,\dots n$. If $u^*_{N-1}(x_{N-1},(N-1)\epsilon)$ is the maximizer
of the right hand term of the above equation, let us denote
\begin{equation}\label{Hdiscrete}
H_{N-1}(x_{N-1},(N-1)\epsilon) ={\cal L}(x_{N-1},
u_{N-1}^*(x_{N-1},(N-1)\epsilon),(N-1)\epsilon)\; .
\end{equation}
In general, for $j=1,\dots,N-1$, the optimal value in (\ref{Dmax})
can be written as
\begin{equation}\label{1}
V_j^* = \max_{\{u_{j}\}} E\left[ {\cal L}(x_j,
u_j,j\epsilon)\epsilon + \sum_{k=1}^{N-j-1} \theta_k
H_{j+k}(x_{j+k},(j+k)\epsilon)\epsilon + \theta_{N-j} V_n^*\right]\; .
\end{equation}
Since
\begin{equation}\label{2}
V_{j+1}^*(x_{(j+1)},(j+1)\epsilon) = E\left[ \sum_{i=0}^{N-j-2}
\theta_i H_{j+i+1}(x_{j+i+1},(j+i+1)\epsilon)\epsilon +
\theta_{N-j-1} V_N^*\right]\; ,
\end{equation}
then, solving $\theta_{N-j-1} V_N^*(x_N)$ in (\ref{2}) and substituting in (\ref{1})
we obtain:
\begin{proposition}\label{T1}
For every initial state $x_0$, the equilibrium value $V^*_j$ of problem
(\ref{Dmax}-\ref{DCE}) can be obtained as the solution of the following
algorithm, which proceeds backward in time from period $N-1$ to period $0$:
\begin{equation}
\theta_{N-j-1}V_j^*(x_j,j\epsilon)  =  \max_{\{u_j\}} E\left[ \theta_{N-j-1} {\cal L}(x_j,
u_j,j\epsilon)\epsilon \right. +\nonumber
\end{equation}
\begin{equation}\label{DDPE1}
+ \sum_{k=1}^{N-j-1} \left(
\theta_{N-j-1}\theta_k - \theta_{N-j}\theta_{k-1}\right)
H_{j+k}(x_{j+k},(j+k)\epsilon)\epsilon +
\end{equation}
\begin{equation}\nonumber
\left.
+ \theta_{N-j}
V_{j+1}^*(x_{j+1},(j+1)\epsilon)\right]\; ,~~~V_N^* = F(x(T))\; ,
\end{equation}
\begin{equation}
x^i_{j+1} = x^i_{j} + f^i(x_{j},u_{j},j\epsilon)\epsilon +
\sum_{l=1}^L\bar{\sigma}^i_l(x_{j},u_{j},j\epsilon)( w^l_{j+1} -
w^l_{j})\; , \label{DDPE3}
\end{equation}
for $i=1,\dots,n$, $j=0,\dots,N-1$. Equations (\ref{DDPE1}-\ref{DDPE3}) are the equilibrium dynamic programming
equations in discrete time, and their solution is the Markov Perfect
Equilibrium (MPE) solution to problem (\ref{Dmax}-\ref{DCE}).
\end{proposition}
In the proof of the proposition we have implicitly assumed that the functions $V_j$
and $V^*_j$ given by (\ref{Dmax}) and (\ref{DDPE1}), respectively, are well defined and finite.

\subsection{The continuous time case: a heuristic approach}

In order to solve Problem (\ref{DE}-\ref{ND}) for a sophisticated agent, first we need to
define what we mean by a Markov equilibrium. Recall that the concept of optimality
plays no role here, since what is optimal for the $t$-agent will not be optimal
(in general) for the $s$-agents, $s>t$. A natural approach to the problem consists in considering first the equilibrium of a sequence of planners
in discrete time (as we have done in the previous section) and then passing to the continuous time limit.
Hence, the equilibrium value function to Problem (\ref{DE}-\ref{ND}) is defined as the
limit when $\epsilon\to 0$ of the discrete stage equilibrium Problem (\ref{Dmax}-\ref{DCE}), and
the Markov Perfect Equilibrium is defined as the solution to the dynamic programming
equation obtained as the (formal) limit when $\epsilon\to 0$ of equations (\ref{DDPE1}-\ref{DDPE3}).
This is probably the most intuitive approach, and was the one used (in a deterministic setting) in Karp (2007). By following these ideas, we will derive a ``modified'' HJB equation in a heuristic way. However, this approach is not rigorous, since the pass to the limit is ``formal'', and needs to be mathematically justified. For a given equilibrium rule $u(x,s)$,
a condition assuring the uniform convergence
(in the mean square sense) of the solution to the discretized equation (\ref{DCE})
to the true solution to (\ref{DE}) is that $f^i$ and $\bar{\sigma}^i_l$ satisfy uniform
growth and Lipschitz conditions in $x$, and are H\"{o}lder continuous of order $1/2$
in the second variable. In the next section we will follow a different and rigorous approach, similar in spirit to the one first suggested in Barro (1999), proving a theorem for the modified HJB equation.

According to the previous definition, let us assume that the equilibrium value function $V^S(x_t,t)$ of the sophisticated $t$-agent, with initial condition $x(t)=x_t$, is of class $C^2$ in $x$, and of class $C^1$ in $t$ (i.e., of class $C^{2,1}$). Since $t=j\epsilon$ and $x^i(t+\epsilon) = x^i(t) + f^i(x(t),u(t),t)\epsilon +
\sum_{l=1}^L\bar{\sigma}^i_l(x(t),u(t),t)( w^l(t+\epsilon) - w^l(t)
)$, then $V^S(x_t,t)=  V_j(x_j,j\epsilon)$ and
\[
V^S(x_{t+\epsilon},t+\epsilon) = V^S(x_t,t) + \left[ V^S_t +
V^S_x\cdot f + \frac{1}{2} tr\left(\bar{\Sigma}\bar{\Sigma}^\prime V^S_{xx}
\right)\right]_{(x_t,u(t),t)}\epsilon +
\]
\[
+ \sum_{i=1}^n
\sum_{l=1}^L\bar{\sigma}^i_l\left.\frac{\partial V^S}{\partial x^i}\right|_{(x_t,t)}
(w^l(t+\epsilon) - w^l(t)) + o(\epsilon)\; .
\]
Since $\theta_k = \exp\left(-\int_0^{k\epsilon} r(s) ds\right)$, then $\theta_{N-j} = \theta_{N-j-1}\left[ 1-r((N-j)\epsilon)\epsilon\right] +
o(\epsilon) = \theta_{N-j-1}\left[ 1-r(T-t)\epsilon \right] +
o(\epsilon)$ and $\theta_{k-1} = \theta_k\left[1+r(k\epsilon)\epsilon\right] +
o(\epsilon)$, and substituting in (\ref{DDPE1}) and simplifying we obtain
\[
0=\max_{\{u(t)\}}E\left[
\left( {\cal L}+V^S_t + V^S_x\cdot f + \frac{1}{2}
tr\left(\bar{\Sigma}\bar{\Sigma}^\prime V^S_{xx}\right) -
r(T-t)V^S(x_t,t) - K\right)\epsilon\right.
\]
\begin{equation}\label{calW}
\left.\left. + \sum_{i=1}^n\sum_{l=1}^L\bar{\sigma}^i_l
\frac{\partial V^S}{\partial x^i}
(w^l(t+\epsilon) - w^l(t)) + o(\epsilon)\right|_{(x_t,u(t),t)}\right]
\end{equation}
where $K(x_t,t)$ is given by
\begin{equation}\label{calK}
K(x_t,t)=E\left[ \sum_{k=1}^{n-j-1} \theta(k\epsilon)\left[
r(k\epsilon) - r(T-t)\right] H_{t+k\epsilon} (x(t+k\epsilon), t+k\epsilon)
\epsilon\right]\; .
\end{equation}
Dividing equation (\ref{calW}) by $\epsilon$, and taking
the limit $\epsilon\to 0$ (and hence $n\to \infty$, since $T=n\epsilon$)
in (\ref{calW}) and (\ref{calK}), we obtain the modified HJB equation
\begin{equation}\label{B1}
r(T-t)V^S+K-\frac{\partial V^S}{\partial t}
=\max_{\{ u\}} \left\{ {\cal L} + V^S_x\cdot f + \frac{1}{2}
tr\left(\bar{\Sigma}\bar{\Sigma}^\prime V^S_{xx}\right)\right\}\; ,
\end{equation}
where
\[
K(x_t,t) = E\left[\int_0^{T-t} \theta(s)\left[ r(s) - r(T-t)\right]
H(x(t+s),t+s) ds \right] =
\]
\[
=E\left[\int_t^{T} \theta(s-t)\left[ r(s-t) - r(T-t)\right]
H(x_s,s) ds\right]\; ,
\]
and $H(x_s,s)={\cal L}(x_s, u^*(x_s,s),s)$. Finally, note that for the equilibrium rule $u^*=u^*(x_s,s)$,
$s\in[t,T]$, by solving the stochastic differential equation
(\ref{DE}) we can
write $x_{t+s}$ as a function of $x_t$ and
$s$ ($x_{t+s}=x(x_t,s)$). Therefore, $H(x_s,s) = H(x_t,s)$ and the Markov Perfect Equilibrium (MPE) $u^*(x,t)$, $t\in[0,T]$, is obtained by solving (\ref{B1}) with
\begin{equation}\label{B2}
V^S(x,T)=F(x)\; ,
\end{equation}
\begin{equation}\label{B3}
K(x,t) = E\left[\int_t^{T} \theta(s-t)\left[ r(s-t) - r(T-t)\right]
H(x,s) ds\right]\; .
\end{equation}

\begin{remark} Assume now that, in addition to the initial
state $x(0)=x_0$, the final state $x(T)=x_T$ is given. In this case,
the terminal condition $W(T,x)=F(x)$ makes no sense ($x(T)$ is fixed).
Instead, we have the extra condition $x(T)=x_T$ in order to
integrate the differential equations.
\end{remark}

\begin{remark} As we have commented previously, the above derivation of the modified HJB equation is heuristic.
The pass to the limit has to be mathematically justified. In stochastic
optimal control, in Fleming and Soner (2006) the convergence of finite
difference approximations to HJB equations is discussed. However, the convergence
is proved by using a method based on viscosity solution techniques, and therefore
it is not applicable to equation (\ref{B1}), which is not a partial differential
equation due to the presence of a non-local term. In fact, for a numerical resolution of the problem in continuous time,
the extension of the results on the convergence of numerical methods to the value function
presented in Kushner and Dupuis (2001) is a topic which deserves attention, although
such extension seems to be not straightforward.
\end{remark}

If there is no final function, equations (\ref{B1}-\ref{B3})
can be written as follows:

\begin{corollary}
If, in Problem (\ref{DE}-\ref{ND}), there is no final function
($F(x(T))=0$), then the modified HJB equation can be written as
\[
\bar{K}-V^S_t = \max_{\{ u\}} \left\{ {\cal L} + V^S_x\cdot f
+\frac{1}{2} tr(\Sigma\Sigma^\prime V^S_{xx})\right\}\; ,
\]
where
\[
\bar{K}(x,t) = E\left[ \int_t^{T} \theta(s-t) r(s-t) H(x,s) ds\right]
\]
and $V^S(x,T)=0$.
\end{corollary}

\noindent {\bf Proof:} Along the equilibrium path,
\[
V^S(x,t)=E\left[\int_t^T\theta(s-t)H(x,s) ds + \theta(T-t)F(x(T))\right]\; .
\]
Therefore
\[
K(x,t) = E\left[\int_t^{T} \theta(s-t)r(s-t) H(x,s) ds - r(T-t)\int_t^T
\theta(s-t)H(x,s) ds\right] =
\]
\[
=\bar{K}(x,t) - r(T-t)V^S(x,t)-\theta(T-t)r(T-t)E\left[F(x(T))\right]\; .
\]
If $F=0$, the result follows by substituting the expression
above in (\ref{B1}).
\hfill $\Box$

\bigskip

If the discount rate is non-constant but the $0$-agent can precommit
at time $t=0$ his or her future behavior, then the corresponding 
(classical) HJB equation characterizing the pre-commitment solution becomes
\begin{equation}\label{commitment}
r(t)V^P-V^P_t = \max_{\{ u\}}
\left\{ {\cal L}+ V^P_x\cdot f +
\frac{1}{2} tr(\Sigma\Sigma^\prime V^P_{xx})\right\}\; .
\end{equation}

When comparing equations (\ref{commitment}) and
(\ref{B1}), there are two differences. First,
the term $r(t)V(x,t)$ in (\ref{commitment}) changes to
$r(T-t)V^S(x,t)$. Second, and more importantly, a new
term $K(x,t)$ appears in equation (\ref{B1}). This new term involves the utility function ${\cal L}$. Note that, in (\ref{B3}), $H(x,s)$ is essentially the function ${\cal L}$ evaluated at the equilibrium rule. This is a substantial change with respect to the standard HJB equation. If the discount rate is constant, $K=0$ and we recover the usual HJB equation. Otherwise, this extra term has to be added, and equation (\ref{B1}) becomes an integro-differential equation. This fact determines a substantial increase in the complexity of the mathematical treatment. Moreover, the modified HJB equation given by (\ref{B1}-\ref{B3}) appears not to be very useful, insofar as it includes implicitly the equilibrium rule (that is, the solution to the problem) in the definition of $K(x,t)$ (via $H(x,s)={\cal L}(x^*_s(x,u^*(x,s),s),u^*(x,s),s)$). In this paper we illustrate how equations (\ref{B1}-\ref{B3}) can be used in order to solve a consumption and portfolio rules problem, by applying a guessing method for searching the solution.

Another fundamental difference between
stochastic optimal control theory and Problem (\ref{DE}-\ref{ND})
comes from the fact that,
since the problem with non-constant discount rate of time preference
is equivalent to a game with a continuous number of agents,
each of whom wants to maximize the expected present discounted value
of current and future welfare, the notion of optimality is
substituted by that of Markov Perfect Equilibrium. If such equilibria
are non unique, the concept of Pareto optimality should be applied
here. Such non-uniqueness of candidate equilibria is usual in an
infinite time setting, and was addressed in Karp (2007) (in a deterministic
infinite horizon context), where a Pareto ranking of steady states
was established. In the consumption and portfolio rules problem
with a finite planning horizon studied in this paper the MPE is unique,
and therefore this problem is avoided.

Things are much easier in the (Mayer) problem, where we are just interested
in maximizing a final expected utility (${\cal L}=0$ in (\ref{ND})).
If the discount rate is constant, it is clear that the optimal
solution is independent of the discount rate.
It is straightforward to see that this property is preserved in the case
of a non-constant discount rate of time preference, not only for
the pre-commitment and naive solutions (where actually we are solving
standard optimal control problems), but also for sophisticated agents.
For instance, let $V$ be the solution when $r=0$. Then $V$ verifies
$-V_t =\max_{\{ u\}} \left\{ V_x\cdot f + \frac{1}{2}tr\left(\Sigma\Sigma^\prime V_{xx}\right)\right\}$.
Now, if $V^S$ is the value function for a sophisticated agent with
(arbitrary) non-constant discounting, from (\ref{B1}) we obtain
($K=0$ in this case) $V^S(x,t)=\theta(T-t) V(x,t)$. Therefore, although
the value function changes in a factor $\theta(T-t)$, the optimal/equilibrium
control-state pair coincides for both problems. Hence, in a pure optimal
portfolio management problem such as Problem (\ref{budget}-\ref{CI})
where we omit consumption ($u(c)=c=0$ in the model), the introduction of
time-inconsistent preferences does not add anything new. However, in
problems where the final time $T$
is a decision variable ($T$ is not prefixed), the changing preferences
of the decision-maker will modify the optimal solution, in general
(the equilibrium final time will be different for naive and sophisticated
agents, and for different discount rates, see Mar\'{\i}n-Solano and Navas
(2009)).

\subsection{The Modified HJB Equation}

In this section we provide a mathematical justification of the dynamic programming equations (\ref{B1}-\ref{B3}). We will follow the underlying idea used (in a deterministic setting) in Barro (1999) for the derivation of the equilibrium rules, consisting in assuming that the decision-maker
at time $t$ can precommit his future behavior during the period $[t,t+\epsilon]$.
In Ekeland and Lazrak (2008) this idea was reformulated by considering that the $t$-agent
is allowed to form a coalition with his immediate successors ($s$-agents, with $s\in[t,t+\epsilon]$), provided that, for $s>t+\epsilon$, the corresponding $s$-agents
choose their equilibrium rule. Then, the equilibrium rule was calculated by taking the
limit $\epsilon\to 0$. It is remarkable that the equilibrium necessary conditions obtained
(in a deterministic setting) in Karp (2007) and Ekeland and Lazrak (2008)) are consistent, although the two approaches are different in nature. Very recently, in Ekeland and Pirvu (2008a, 2008b), the approach in Ekeland and Lazrak (2008) has been extended to a stochastic setting, and applied to some portfolio management problems, providing an alternative approach to ours. We refer the reader to these papers for a formal presentation and a mathematically rigorous analysis of the non-standard optimal control problem with non-constant discounting of time preference.

Let us assume that the $t$-agent can precommit his future behavior during
the period $[t,t+\epsilon]$. Then, given an equilibrium rule $u(x,s)$, for
$s>t+\epsilon$,
\[
V^S(x_t,t)=\max_{\{ u(t)\}} E_{x_t,t}\left[\int_t^T\theta(s-t){\cal L}(x(s),u(s),s) ds + \theta(T-t)F(x(T))\right]
=
\]
\begin{equation}\label{new1}
= \max_{\{ u(t)\}}E_{x_t,t}\left\{\int_t^{t+\epsilon}\theta(s-t){\cal L}(x(s),u(s),s) ds + \right.
\end{equation}
\[
+\left.E_{x_{t+\epsilon},t+\epsilon}\left[\int_{t+\epsilon}^T\theta(s-t)H(x(s),s) ds + \theta(T-t)F(x(T))\right]
\right\}\; ,
\]
where $H(x(s),s)={\cal L}(x(s),u^*(x(s),s),s)$, with $u^*(x,s)$ the equilibrium rule, is defined as in the previous section. Note that, due to the non-constant discounting, we cannot write the equation above in terms of $V^S(t+\epsilon,x(t+\epsilon))$ in the usual form. Instead,
\begin{equation}\label{new2}
V^S(x_{t+\epsilon},t+\epsilon)=
\end{equation}
\[
=\max E_{t+\epsilon,x_{t+\epsilon}}\left[\int_{t+\epsilon}^T\theta(s-t-\epsilon) H(x(s),s) ds + \theta(T-t-\epsilon)F(x(T))\right]\; .
\]
Let $\bar{u}$ be the maximum in (\ref{new1}). Then, by solving $E_{t+\epsilon,x_{t+\epsilon}}F(x(T))$ in (\ref{new1}) and (\ref{new2}) and identifying terms we obtain
\[
\theta(T-t-\epsilon)\left[V^S(x,t)-E_{x_t,t}\int_t^{t+\epsilon}\theta(s-t){\cal L}(x(s),\bar{u},s) ds - \right.
\]
\[
\left. - E_{x_{t+\epsilon},t+\epsilon}\int_{t+\epsilon}^T\theta(s-t)H(x(s),s) ds\right] =
\]
\[
= \theta(T-t)\left[V^S(x(t+\epsilon),t+\epsilon)- E_{x_{t+\epsilon},t+\epsilon}\int_{t+\epsilon}^T\theta(s-t)H(x(s),s) ds\right]\; .
\]
If $V^S(x,t)$ is of class $C^{2,1}$, applying the Ito rule to $V^S(x(t+\epsilon),t+\epsilon)$, dividing by $\epsilon$ and taking the limit $\epsilon\to 0$ we recover the dynamic programming equations (\ref{B1}-\ref{B3}).

Next, we make rigorous the previous reasoning by proving a theorem, which is an extension of an standard result in stochastic optimal control theory (see, e.g., Fleming and Rishel (1975) or Fleming and Soner (2006)). Let
\[
A^v=\frac{\partial}{\partial t} + \sum_{i=1}^n f^i\frac{\partial}{\partial x^i} + \sum_{i,j=1}^n \sigma_{ij}\frac{\partial^2}{\partial x^i\partial x^j}
\]
be the backward evolution operator, with $\sigma_{ij} = \sum_{l=1}^L\bar{\sigma}^i_l\bar{\sigma}^j_l$, as usual.

Given initial data $(t,x)$, we call $\pi=(\Omega,\{{\cal F}_s\},P,s(\cdot),u(\cdot))$ an {\sl admissible control system} if $(\Omega,{\cal F}_T,P)$ is a probability space, $\{{\cal F}_s\}$ is an increasing family of $\sigma$-algebras ($t\leq s\leq T$), and $x(\cdot)$, $u(\cdot)$ are stochastic processes on $[t,T]$ such that:
\begin{enumerate}
\item $x(s)\in X$ for $s\in[t,T]$, $x(t)=x$, the sample paths $x(\cdot,\omega)$ are right continuous and have left hand limits, and $x(s)$ is ${\cal F}_s$-measurable;
\item $u(s)\in U$ for $s\in[t,T]$, $u(s)$ is ${\cal F}_s$-measurable and $u(\cdot,\cdot)$ is measurable;
\item For all function $\Phi(t,x)$ of class $C^{1,2}$ with polynomial growth of $\Phi$ and $A^v\Phi$, satisfying $E_{t,x}|\Phi(T,x(T))|<\infty$, and $E_{t,x}\int_t^T|A^{u(s)}\Phi(s,x(s))|ds<\infty$, the Dynkin formula holds:
\[
E_{t,x}\Phi(T,x(T))-\Phi(t,x)=E_{t,x}\int_t^T A^{u(s)}\Phi(s,x(s))\, ds\; .
\]
\end{enumerate}
Then we have:
\begin{theorem}\label{teoremanuevo}
Let $V^S(t,x)$ be a function of class $C^{1,2}$ with polynomial growth of $V^S$ and $A^v V^S$, solution to (\ref{B1}-\ref{B3}). If there exists an admissible system $\pi^*=(\Omega^*,\{{\cal F^*}_s\},x^*(\cdot),u^*(\cdot))$ such that $u^*(s)$ solves the right hand term in equation (\ref{B1}) for Lebesgue$\times P^*$-almost all $(s,\omega)\in[t,T]\times\Omega^*$, then
\[
V^S(t,x) = E_{t,x}\left\{\int_t^T \theta(s-t){\cal L}(s,x(s),u^*(s))\, ds + \theta(T-t)F(x(T))\right\}\; .
\]
\end{theorem}

\noindent {\bf Proof:} For an admissible control system $\pi^*$ satisfying the conditions in the Theorem, since $u^*(s)\in U$, from (\ref{B1}) we have
\[
- A^{u^*(s)} V^S(t,x(t)) = {\cal L}(t,x(t),u^*(t)) - r(T-t)V^S(t,x(t)) - K(t,x(t)) \; .
\]
From the Dynkin formula and the above expression we obtain
\begin{equation}\label{new5}
V^S(t,x) = E_{t,x}\left[ \int_t^T-A^{u^*(s)} V^S(s,x(s))\, ds + F(x(T))\right] =
\end{equation}
\[
= E_{t,x}\left[ \int_t^T\left({\cal L}(s,x(s),u^*(s)) - r(T-s) V^S(s,x(s)) - K(s,x(s))\right) \, ds + F(x(T))\right]\; ,
\]
where
\[
E_{t,x}\left\{\int_t^T K(s,x(s))\, ds\right\} = 
\]
\[
= E_{t,x}\left\{\int_t^T ds\,\left[\int_s^T d\tau\,\theta(\tau-s)[r(\tau-s) - r(T-s)] {\cal L}(\tau,x(\tau),u^*(\tau))\right]\right\}\; .
\]
Next, note that
\[
E_{t,x}\left\{\int_t^T ds\, r(T-s)\left[\int_s^T d\tau\,\theta(\tau-s) {\cal L}(\tau,x(\tau),u^*(\tau))\right]\right\} =
\]
\[
= E_{t,x}\left\{\int_t^T ds\, r(T-s) V^S(s,x) - \int_t^T ds\, \theta(T-s) r(T-s) F(x(T))\right\} =
\]
\[
= E_{t,x}\left\{\int_t^T ds\, r(T-s) V^S(s,x) - F(x(T)) + \theta(T-t) F(x(T))\right\}\; .
\]
By substituting in (\ref{new5}) and simplifying we obtain
\begin{equation}\label{new6}
V^S(t,x)=E_{t,x}\left\{ \int_t^T{\cal L}(s,x(s),u^*(s))\, ds -\right.
\end{equation}
\[
\left. - \int_t^T ds\left[ \int_s^T d\tau \theta(\tau-s) r(\tau-s) {\cal L}(\tau,x(\tau),u^*(\tau)) \right] + \theta(T-t) F(x(T))  \right\}\; .
\]
Finally, note that
\[
E_{t,x}\left\{ \int_t^T ds\,\left[ \int_s^T d\tau\, \theta(\tau-s) r(\tau-s) {\cal L}(\tau,x(\tau),u^*(\tau))\right] \right\} =
\]
\[
= E_{t,x}\left\{ \int_t^T d\tau\, {\cal L}(\tau,x(\tau),u^*(\tau)) \left[ \int_t^\tau ds\, \theta(\tau-s) r(\tau-s) \right] \right\}
=
\]
\[
=
E_{t,x}\left\{ \int_t^T d\tau\, (1-\theta(\tau-t)){\cal L}(\tau,x(\tau),u^*(\tau)) \right\}\; .
\]
Then the result follows by substituting the expression above in (\ref{new6}).
\hfill $\Box$

In the consumption and portfolio rules problem analyzed in this paper, the stochastic differential equations are linear (in fact, the equilibrium control rules are also linear in the state variable), and satisfy the hypothesis for the existence and uniqueness of solutions and for the Dynkin formula. For more details on admissible controls and conditions under which the Dynkin formula holds, see, e.g., Fleming and Soner (2006) or Fleming and Rishel (1975).

\section{Optimal Portfolios for Time-Inconsistent\\ Investors}

\subsection{General Setting}

In this section, we analyze the consequences of introducing
a non-constant discount rate in time preference into the classical
solution by Merton (1969, 1971) for the optimal consumption and portfolio
problem. Let us briefly describe the basic parameters of the problem.

The standard Ito processes model for a financial market consists of
(m+1) securities. One of them is risk-free (a cash account, for instance), and
the price $P_0(t)$ of 1 unit is assumed to evolve according to the
ordinary differential equation $\frac{d P_0(t)}{P_0(t)} = \mu_0 d t$,
where $\mu_0>0$ and $P_0(0)=p_0>0$. There are also $m$
risky assets (stocks, for instance), whose prices
$P_i(t)$, $i=1,\dots m$, evolve according to a geometric Brownian
motion stochastic process:
\begin{equation}\label{risky}
\frac{d P_i(t)}{P_i(t)} = \mu_i d t +
\sum_{k=1}^{L}\bar{\sigma}_{ik} d \bar{z}_k(t)\; ,~~~i=1,\dots m\; ,
\end{equation}
where $P_i(0)=p_i>0$, $(\bar{z}_1(t),\dots,\bar{z}_L(t))$ is an $L$-dimensional
standard Brownian motion process, and $\bar{z}_k(t)$ are mutually
independent Brownian motions. For the sake of simplicity, we will
assume that $\mu_0$ and the drift vector of the risky assets ${\bf{\mu}}
=(\mu_1,\dots,\mu_m)$ are constant.

From the diffusion matrix $\bar{\Sigma} = \left(
\bar{\sigma}_{ik}\right)$, $i=1,\dots,m$, $k=1,\dots,l$, we can
define the variance-covariance matrix $\Sigma =
\bar{\Sigma}\cdot\bar{\Sigma}^\prime = \left( \sigma_{ij}\right)$,
$i,j=1,\dots m$, whose coefficients are given by $\sigma_{ij} =
\sum_{k=1}^l \bar{\sigma}_{ik}\bar{\sigma}_{jk}$. Note that
$\Sigma$ is symmetric ($\sigma_{ij} = \sigma_{ji}$). We will
assume that $\Sigma$ is positive definite. In particular,
this implies that $\sigma_{ii}>0$ (all $m$ risky assets are indeed
risky) and $\Sigma$ is nonsingular ($\det \Sigma>0$).
Elements $\sigma_{ii}$ are usually denoted by $\sigma_i^2$,
hence $\sigma_i = (\sigma_{ii})^{1/2}$.

By defining $z_i(t)= \frac{1}{\sigma_i}\sum_{k=1}^l
\bar{\sigma}_{ik} \bar{z}_k(t)$, which are correlated standard Brownian
motions with $\hbox{Cov}\, (z_i(t),z_j(t)) =
\frac{\sigma_{ij}}{\sigma_i\sigma_j} t$, Equation
(\ref{risky}) becomes
\[
d P_i = \mu_i P_i d t + \sigma_i P_i d z_i\;
,~~~i=1,\dots m\; ,
\]
with $dz_i dz_j=\rho_{ij}dt$ for $i,j=1,\dots m$, where
$\rho_{ij}=\frac{\sigma_{ij}}{\sigma_i\sigma_j}$.
Therefore, the problem for the $t$-agent consists
in solving (\ref{CI}) subject to (\ref{budget}).

In terms of the Wiener $L$-dimensional process with independent
components $(\bar{z}_1,\dots,\bar{z}_L)$, the budget equation
is
\begin{equation}\label{budget2}
d W = \left[\sum_{i=1}^{m} w_i(\mu_i-\mu_0)W+(\mu_0 W-c)\right] ds
+ \sum_{i=1}^{m}\sum_{k=1}^L w_i\bar{\sigma}_{ik} W d\bar{z}_k\; .
\end{equation}

Let us briefly recall the solution
with a constant discount rate $\rho$. In this case the agent must solve
the HJB equation
\[
\rho V-V_t = \max_{\{ c,w_i\}} \left\{ u(c) +
\left[\sum_{j=1}^{m} w_j(\mu_j-\mu_0)W+(\mu_0 W-c)\right]
V_W +\right.
\]
\begin{equation}\label{estandar}
+\left.\frac{1}{2}\sum_{j=1}^m\sum_{k=1}^m
w_j w_k \sigma_{jk} W^2 V_{WW}\right\}\; .
\end{equation}
By solving the maximization problem in $w_i$, $i=1,\dots,m$,
we obtain the standard optimal portfolio rule
\begin{equation}\label{portfolio}
{\bf w}=-\frac{V_W}{WV_{WW}}\Sigma^{-1}({\bf{\mu}}-\mu_0\cdot{\bf 1})\; ,
\end{equation}
where ${\bf{\mu}}=(\mu_1,\dots,\mu_m)$ and ${\bf 1}=(1,\dots,1)$.
As for the optimal consumption, from the maximization problem in
$c$ in Equation (\ref{estandar}) we obtain
\begin{equation}\label{consumo}
u^\prime(c)=V_W\; .
\end{equation}

Next, let us assume that the discount rate $r(t)$ of time preference is
non-constant. For the general case, let us describe the so-called
pre-commitment solution, and the solution for naive and sophisticated
agents.

\bigskip

\noindent {\bf Pre-commitment Solution:} If the $0$-agent can
precommit his future behavior, he must solve
the corresponding HJB equation
\[
r(t)V^P-V^P_t = \max_{\{ c,w_i\}} \left\{ u(c) +
\left[\sum_{j=1}^{m} w_j(\mu_j-\mu_0)W+(\mu_0 W-c)\right]
V^P_W +\right.
\]
\begin{equation}\label{compromiso}
+\left.\frac{1}{2}\sum_{j=1}^m\sum_{k=1}^m
w_j w_k \sigma_{jk} W^2 V^P_{WW}\right\}\; .
\end{equation}
Since the right hand term in (\ref{compromiso}) coincides with that in
(\ref{estandar}), then the optimal consumption and portfolio
rules are given by (\ref{consumo}) and (\ref{portfolio}), respectively,
with $V$ replaced by $V^P$.

\bigskip

\noindent {\bf Solution for a Naive Agent:} Naive $t$-agents will
solve the problem by looking for the solution to the HJB equation
\[
r(\tau-t)V^N-V^N_\tau = \max_{\{ c,w_i\}} \left\{ u(c) +
\left[\sum_{j=1}^{m} w_j(\mu_j-\mu_0)W+(\mu_0 W-c)\right]
V^N_W +\right.
\]
\begin{equation}\label{naive}
+\left.\frac{1}{2}\sum_{j=1}^m\sum_{k=1}^m
w_j w_k \sigma_{jk} W^2 V^N_{WW}\right\}
\end{equation}
where the value function for the naive $t$-agent is $V^N(W,\tau)$,
for $\tau\in[t,T]$. From the maximization problem in (\ref{naive})
we again obtain that the optimal consumption and portfolio
rules are given by (\ref{consumo}) and (\ref{portfolio}), with $V$
replaced by $V^N$. In order to construct the actual trajectory, we will patch together the solutions $u^t(t)$.

\bigskip

\noindent {\bf Solution for a Sophisticated Agent:}
From Theorem 1, in order to solve Problem
(\ref{CI}) subject to (\ref{budget2}), we analyze the modified HJB equation
(\ref{B1}), which for our particular problem becomes
\[
r(T-t)V^S+K-V^S_t = \max_{\{ c,w_i\}} \left\{ u(c) +
\left[\sum_{j=1}^{m} w_j(\mu_j-\mu_0)W+(\mu_0 W-c)\right]
V^S_W + \right.
\]
\begin{equation}\label{HJBCI}
\left. +\frac{1}{2}\sum_{j=1}^m\sum_{k=1}^m
w_j w_k \sigma_{jk} W^2 V^S_{WW}\right\}\; ,
\end{equation}
where $K$ is given by (\ref{B3}). Once again, the optimal consumption and portfolio
rules are given by (\ref{consumo}) and (\ref{portfolio}), with $V$
replaced by $V^S$.

\bigskip

From Equation (\ref{portfolio}) it becomes clear that, if for every
non-constant discount rate of time preference $r(s)$,
$\frac{V^S_W}{WV^S_{WW}}$ is constant, then
the investment strategy will be independent of $r(s)$ and it
will be observationally equivalent to the constant discount rate
case. From the solution to the problem
in Merton (1969, 1971), natural candidates for this observational
equivalence are the CRRA (constant relative risk averse) utility
functions, namely the logarithmic and potential functions. In the
remaining subsections of the paper, we show how
this observational equivalence exists in the investment strategy (not
in the consumption rule) for the logarithmic and potential utility
functions, but not for more general utility functions, such as
the exponential CARA (constant absolute risk averse) utility
function.

\subsection{Logarithmic Utility Function}

First of all, let us analyze the log-utility case,
$u(c)=\hbox{ln}\, c$, with final function
$F(W(T))=a\, \hbox{ln}\, (W(T))$.

In the case of a constant discount rate $\rho$, the agent must solve
the HJB equation (\ref{estandar}). From a symmetry argument (see, for instance, Boyd (1990) and Chang (2004),
pp. 193-194)
it can be proved that $V(W,t)=\alpha(t)\,\hbox{ln}\, W+\beta(t)$.
In fact this symmetry argument can be applied to all the solutions in this section. From (\ref{consumo}) we obtain
$c= \left(V_W\right)^{-1}=W/\alpha(t)$ and, by substituting in (\ref{portfolio}) the
optimal portfolio rule becomes
\begin{equation}\label{portfoliostandard}
{\bf w}=\Sigma^{-1}({\bf{\mu}}-\mu_0\cdot{\bf 1})\; .
\end{equation}
By substituting in (\ref{estandar}), the choice of the value function
proves to be consistent and the optimal consumption rule is determined
by
\begin{equation}\label{consumostandard}
c(t)=\frac{\rho W_t}{1-\left(1- a\rho\right) e^{-\rho(T-t)}}\; .
\end{equation}

Next, for the general case of non-constant discounting,
we solve and compare the solutions for pre-commitment, naive and
sophisticated agents.

\bigskip

\noindent {\bf Pre-commitment Solution:} We must solve equation
(\ref{compromiso}). Once again, we know that the solution will be of the form
$V^P(W,t)=\alpha^P(t)\,\hbox{ln}\, W+\beta^P(t)$. Then, the optimal
consumption and portfolio rules are given by
$c=W/\alpha^P(t)$ and (\ref{portfoliostandard}),
respectively. By substituting in (\ref{compromiso}), we
obtain that $\alpha^P(t), \beta^P(t)$ are the solution to the first
order linear differential equation system
\begin{equation}\label{logI1}
\dot{\alpha}^P-r(t)\alpha^P+1=0\; ,
\end{equation}
\[
\dot{\beta^P}-r(t)\beta^P+\left[\frac{1}{2}({\bf{\mu}}-
\mu_0\cdot{\bf 1})^\prime\Sigma^{-1}({\bf{\mu}}-\mu_0\cdot{\bf 1}) +
\mu_0\right]\alpha^P - \hbox{ln}\,\alpha^P-1=0\; ,
\]
with $\alpha^P(T)=a$, $\beta^P(T)=0$. By solving $\alpha^P(t)$
we obtain
\begin{equation}\label{logcompromiso}
c^{P}(t)=\frac{\theta(t) W_t}{a\theta(T) +\int_t^T\theta(s)\, ds}\; .
\end{equation}


\noindent {\bf Solution for a Naive Agent:} Naive $t$-agents will solve
the problem by looking for the solution of the HJB equation (\ref{naive}).
By guessing $V^N(W,\tau)=\bar{\alpha}^N(\tau)\,\hbox{ln}\, W+\bar{\beta}^N(\tau)$,
we obtain (\ref{portfoliostandard}) and
$c(\tau)=W/\bar{\alpha}^N(\tau)$ where $\bar{\alpha}^N(\tau)$
is the solution to the first order linear differential equation
$\dot{\bar{\alpha}}^N-r(\tau-t)\bar{\alpha}^N+1=0$, $\bar{\alpha}^P(T)=a$,
which is given by
\begin{equation}\label{logcasinaive}
\bar{\alpha}^N(\tau)=\frac{1}{\theta(\tau-t)}\left[ a\theta(T-t)
+\int_\tau^T\theta(s-t)\,ds\right]\; .
\end{equation}
Since the $t$-agent will not be time consistent for $\tau>t$, the
actual consumption rule is obtained from the equation above for the case $\tau=t$, and therefore
\begin{equation}\label{lognaive}
c^N(t)=\frac{W_t}{a\theta(T-t) +\int_t^T\theta(s-t)\,ds}\; .
\end{equation}


\noindent {\bf Solution for a Sophisticated Agent:}
A sophisticated agent will look for the solution of the modified
HJB equation (\ref{HJBCI}), with $K$ given by (\ref{B3}). From the maximization
problem, the equilibrium consumption and portfolio rules are
$c^S(t)=1/V^S_W$ and (\ref{portfoliostandard}). Since the value function satisfies
the same symmetry as in the previous problems, it will be necessarily of the form $V^S(W,t)=\alpha^S(t)\,\hbox{ln}\,
W+\beta^S(t)$. Let us verify the integro-differential equation (\ref{HJBCI}) by the candidate solution. If the choice proves to be consistent, then
$c^S(t)=W/\alpha^S(t)$.
The solution to the stochastic differential equation (\ref{budget2}) is
$W(s)=W_t\exp[\Lambda_t(s)]$, where
\[
\Lambda_t(s)= \left[\mu_0+\sum_{i=1}^m w_i(\mu_i-\mu_0)-\frac{1}{2}\sum_{i=1}^m\left(\sum_{k=1}^L w_i \bar{\sigma}_{ik}\right)^2\right](s-t) -
\]
\[
-\int_t^s \frac{d\tau}{\alpha^S(\tau)} + \sum_{i=1}^m\sum_{k=1}^L w_i\bar{\sigma}_{ik} [w_k(s)-w_k(t)]\; .
\]
Then,
\[
K=E\left[ \int_t^T\theta(s-t)\left[ r(s-t)-r(T-t)\right]\hbox{ln}\, \frac{W(s)}{\alpha^S(s)} \, ds\right] =
\]
\[
= E\left[ \int_t^T\theta(s-t)\left[ r(s-t)-r(T-t)\right]\left[ \hbox{ln}\, W_t + \Lambda_t(s) - \hbox{ln}\,
\alpha^S(s) \right]\, ds\right]\; .
\]
By substituting in (\ref{HJBCI}) and simplifying we obtain
\[
\left[\int_t^T \theta(s-t)[r(s-t)-r(T-t)]\, ds -
\dot{\alpha}^S(t)+r(T-t)\alpha^S(t)-1\right] \hbox{ln}\, W_t =
\]
\[
=- \int_t^T \theta(s-t)[r(s-t)-r(T-t)] \left[\Lambda_t(s)-\hbox{ln}\,\alpha^S(s)\right]\, ds - r(T-t)\beta^S(t) +
\]
\[
+ \dot{\beta}^S(t) - \hbox{ln}\,\alpha^S(t) + \left[\mu_0+\sum_{i=1}^m w_i(\mu_i-\mu_0)-\frac{1}{2}\sum_{j,k=1}^m w_j w_k \sigma_{jk}\right]\alpha^S(t) - 1\; .
\]
Since the equation above must be satisfied for every $W_t$, then
necessarily
\begin{equation}\label{logI2}
\dot{\alpha}^S-r(T-t)\alpha^S+1=\int_t^T\theta(s-t)[r(s-t)-r(T-t)]\, ds\; .
\end{equation}
Here, we can compare equation (\ref{logI1}) describing the precommitment solution, and equation (\ref{logI2}) for the sophisticated agent. If the discount rate is constant, then $r(t)=r(T-t)=\rho$ and the integral term in (\ref{logI2}) vanishes. Otherwise, it contributes to the solution. Using that $\int_t^T\theta(s-t) r(s-t)\, ds = \left.-\theta(s-t)
\right|_t^T = -\theta(T-t)+1$ we obtain
\[
\dot{\alpha}^S-r(T-t)\alpha^S=-\theta(T-t)-r(T-t)\int_t^T\theta(s-t)\, ds\; .
\]
The general solution to this first order linear differential equation, with the boundary condition $\alpha^S(T)=a$, is
$\alpha^S(t)= a\theta(T-t) +\int_t^T\theta(s-t)\,ds$. Hence,
\begin{equation}\label{logsophisticated}
c^S(t)=\frac{W_t}{a\theta(T-t) +\int_t^T\theta(s-t)\,ds}\, ,
\end{equation}
which coincides with the solution obtained for a naive agent. Of course,
this is a special feature of the logarithmic utility function, as we show next.

\subsection{Potential Utility Function}

Next, let us study the problem with a (isoelastic) potential
utility function, $u(c)=c^\gamma/\gamma$,
$\gamma<1$, $\gamma\neq 0$,
with final function $F(W(T))=a[W(T)]^\gamma/\gamma$.

As above, first we recall the solution with a constant discount rate $\rho$.
From the right hand term in (\ref{estandar}) we obtain (\ref{consumo})
and (\ref{portfolio}). As a candidate to the value function we guess
$V(W,t)=\alpha(t) [W(t)]^\gamma/\gamma$. Once again, this
choice is justified in Boyd (1990) from a symmetry argument, which is also
applied to the pre-commitment, naive and sophisticated solutions.
From (\ref{consumo}) we obtain $c= (\alpha(t))^{-\frac{1}{1-\gamma}} W$
and, from (\ref{portfolio}), the
optimal portfolio rule becomes
\begin{equation}\label{portfoliostandardp}
{\bf w}=\frac{1}{1-\gamma}\Sigma^{-1}({\bf{\mu}}-\mu_0\cdot{\bf 1})\; .
\end{equation}
By substituting in (\ref{estandar}) we get a Bernoulli equation, and solving it we obtain
\begin{equation}\label{consumostandardp}
c(t)=\frac{\left(\rho-\delta^p\right)e^{\frac{\rho-\delta^p}{1-\gamma}(T-t)}W_t}{a\left(\rho-\delta^p\right)+(1-\gamma)\left( e^{\frac{\rho-\delta^p}{1-\gamma}(T-t)} - 1\right)}\; ,
\end{equation}
where
\begin{equation}\label{alphapp}
\delta^p=\mu_0\gamma+\frac{1}{2}\frac{\gamma}{1-\gamma}
\left({\bf{\mu}}-\mu_0\cdot{\bf 1}\right)^\prime\Sigma^{-1}
\left({\bf{\mu}}-\mu_0\cdot{\bf 1}\right)\; .
\end{equation}

Next, we assume that the discount rate of time preference is non-constant.

\bigskip

\noindent {\bf Pre-commitment Solution:} We guess $V^P(W,t)=\alpha^P(t)[W(t)]^\gamma/\gamma$ (\ref{compromiso}). Then, the equilibrium consumption and
portfolio rules  are given by $c= (\alpha^P(t))^{-\frac{1}{1-\gamma}} W$ and
(\ref{portfoliostandardp}), respectively. By substituting in (\ref{compromiso})
we obtain that $\alpha^P(t)$ is the solution to the Bernoulli equation
$\dot{\alpha}^P=(r(t)-\delta^p)\alpha^P-(1-\gamma)(\alpha^P)^{-\frac{\gamma}{1-\gamma}}$,
$\alpha^P(T)=a$,
where $\delta^p$ is given by (\ref{alphapp}). By solving it we obtain
\begin{equation}\label{pcompromiso}
c^P(t)=\frac{\left(\frac{\theta(t)}{\theta(T)}e^{-\delta^p(T-t)}\right)^{\frac{1}{1-\gamma}} W_t}{\left( a+
\int_t^T \left(\frac{\theta(s)}{\theta(T)}e^{-\delta^p(T-s)}\right)^{\frac{1}{1-\gamma}}
\,
ds\right)}\; .
\end{equation}


\noindent {\bf Solution for a Naive Agent:} We must solve the HJB equation (\ref{naive}).
By guessing $V^N(W,\tau)=\bar{\alpha}^N(\tau) [W(\tau)]^\gamma/\gamma$ and
substituting in (\ref{naive}) we obtain (\ref{portfoliostandardp}) and
$c(\tau)= (\bar{\alpha}^N(\tau))^{-\frac{1}{1-\gamma}} W$, where
$\bar{\alpha}^N(\tau)$ is the solution to
$\dot{\bar{\alpha}}^N=(r(\tau-t)-\delta^p)\bar{\alpha}^N-(1-\gamma)
(\bar{\alpha}^N)^{-\frac{\gamma}{1-\gamma}}$, $\bar{\alpha}^N(T)=a$,
which is given by
\[
\bar{\alpha}^N(\tau)=e^{\delta^p(T-\tau)}\frac{\theta(T-t)}{\theta(\tau-t)} \left[ a+
\int_\tau^T \left(\frac{\theta(s-t)}{\theta(T-t)}e^{-\delta^p(T-s)}\right)^{\frac{1}{1-\gamma}}
\,
ds\right]^{1-\gamma}\; .
\]
The actual consumption rule, which is obtained for $\tau=t$, is
determined by
\begin{equation}\label{pnaive}
c^N(t)=\frac{e^{-\frac{\delta^p(T-t)}{1-\gamma}} W}{\left[\theta(T-t)\right]^{\frac{1}{1-\gamma}}\left(a+
\int_t^T \left(\frac{\theta(s-t)}{\theta(T-t)}e^{-\delta^p(T-s)}\right)^{\frac{1}{1-\gamma}}
\,
ds\right)}\; .
\end{equation}


\noindent {\bf Solution for a Sophisticated Agent:}
Let us look for the solution of the modified
HJB equation (\ref{HJBCI}), with $K$ given by (\ref{B3}). Once again,
we guess $V^S(W,t)=\alpha^S(t) [W(t)]^\gamma/\gamma$ and verify the integro-differential equation (\ref{HJBCI}) for this solution. By substituting in (\ref{HJBCI}) we obtain
(\ref{portfoliostandardp}) and $c(t)=
(\alpha^S(t))^{-\frac{1}{1-\gamma}} W$, where $\alpha^S(t)$ is the solution
to the integro-differential equation
\begin{equation}\label{sofpot}
\dot{\alpha}^S=(r(T-t)-\delta^p)\alpha^S-(1-\gamma)
(\alpha^S)^{-\frac{\gamma}{1-\gamma}} +
\end{equation}
\[
+ \int_t^T\theta(s-t)[r(s-t)-r(T-t)]
(\alpha^S(s))^{-\frac{\gamma}{1-\gamma}} e^{\gamma\int_t^s\Delta(\tau) d\tau}
\, ds\; ,
\]
where $\Delta(\tau)=\mu_0 + \frac{1}{1-\gamma}\left({\bf{\mu}}-\mu_0\cdot{\bf 1}\right)^\prime\Sigma^{-1}
\left({\bf{\mu}}-\mu_0\cdot{\bf 1}\right) - (\alpha^S(\tau))^{-\frac{1}{1-\gamma}}$. In comparison with the Bernoulli equation describing the precommitment solution, in (\ref{sofpot}) $r(t)$ is replaced by $r(T-t)$, and a new integral term appears, turning the Bernoulli equation into a very complicated highly non-linear integro-differential equation.

\subsection{Exponential Utility Function}

Finally, let us solve the problem for the constant absolute
risk aversion utility function $u(c)=
-e^{-\gamma c}/\gamma$, $\gamma>0$,
with final function $F(W(T))=-a e^{-\gamma W}$.

In the constant discount rate case, we guess $V(W,t)=
-a e^{-\gamma(\alpha(t)+\beta(t)W)}$ with
$\alpha(T)=0$, $\beta(T)=1$ (Boyd (1990), Chang (2004),
pp. 193-194), and once again we can replicate the same symmetry argument for the pre-commitment,
naive and sophisticated solutions. We proceed as before to obtain
\begin{equation}\label{consumoestandardexp}
c= \alpha(t)+\beta(t)W-\frac{\hbox{ln}\,(a\gamma\beta(t))}{\gamma}
\end{equation}
and
\begin{equation}\label{portfoliostandardexp}
{\bf w}=\frac{1}{\gamma\beta(t)W}\Sigma^{-1}({\bf{\mu}}-\mu_0\cdot{\bf 1})\; .
\end{equation}
By substituting in (\ref{estandar}), after several calculations we obtain
\begin{equation}\label{betastandardexp}
\beta(t)=\frac{\mu_0}{1+\left(\mu_0-1\right) e^{-\mu_0(T-t)}}\; ,
\end{equation}
\begin{equation}\label{alphastandardexp}
\alpha(t)=-\frac{1}{\gamma}e^{-\int_t^T\beta(s)\, ds}\int_t^T \left[\delta^e(s)
-\rho\right] e^{\int_s^T\beta(\tau)\, d\tau}\, ds\; ,
\end{equation}
where $\delta^e(t) = \beta(t)-\frac{1}{2}({\bf{\mu}}-\mu_0\cdot{\bf 1})^\prime
\Sigma^{-1}({\bf{\mu}}-\mu_0\cdot{\bf 1})-\beta(t)\hbox{ln}\,(a\gamma
\beta(t))$.

If the case of non-constant discounting we obtain the following solutions:

\bigskip

\noindent {\bf Pre-commitment Solution:} By guessing
$V^P(W,t)= -a e^{-\gamma(\alpha^P(t)+\beta^P(t)W)}$,
the associated equilibrium consumption and portfolio rules are
given by (\ref{consumoestandardexp}-\ref{portfoliostandardexp}), with $\alpha(t),\beta(t)$
replaced by $\alpha^P(t),\beta^P(t)$. Moreover, $\beta^P(t)=\beta(t)$, and
\begin{equation}\label{alphaprecomexp}
\alpha^P(t)=-\frac{1}{\gamma}e^{-\int_t^T\beta(s)\, ds}\int_t^T \left[\delta^e(s)
-r(s)\right] e^{\int_s^T\beta(\tau)\, d\tau}\, ds\; .
\end{equation}


\noindent {\bf Solution for a Naive Agent:} We guess
$V^N(W,\tau)= -a e^{-\gamma(\bar{\alpha}^N(\tau)+\bar{\beta}^N(\tau)W)}$
in equation (\ref{naive}).
As above, the consumption and portfolio rules coincide with those
in (\ref{consumoestandardexp}-\ref{portfoliostandardexp}), with $\alpha(t),\beta(t)$
replaced by $\alpha^N(t),\beta^N(t)$. Once again, $\beta^N(t)=\beta(t)$. Since
$\bar{\alpha}^N(\tau)=-\frac{1}{\gamma}e^{-\int_\tau^T\beta(s)\, ds}\int_\tau^T \left[
\delta^e(s) -r(s-t)\right] e^{\int_{s}^T\beta(\bar{s})\, d\bar{s}}\, ds$,
taking $\tau=t$ we obtain
\begin{equation}\label{alphanaiveexp}
\alpha^N(t)=-\frac{1}{\gamma}e^{-\int_t^T\beta(s)\, ds}\int_t^T \left[
\delta^e(s) -r(s-t)\right] e^{\int_{s}^T\beta(\tau)\, d\tau}\, ds\; .
\end{equation}


\noindent {\bf Solution for a Sophisticated Agent:} In order to solve the modified
HJB equation (\ref{HJBCI}), with $K$ given by (\ref{B3}),
we guess $V^S(W,t)= -a e^{-\gamma(\alpha^S(t)+\beta^S(t)W)}$. Let us verify the
integro-differential equation (\ref{HJBCI}) for this solution.
The consumption and portfolio rules are given by
(\ref{consumoestandardexp}-\ref{portfoliostandardexp}), with $\alpha(t),\beta(t)$
replaced by $\alpha^S(t),\beta^S(t)$. By substituting in (\ref{HJBCI}) we obtain
\[
-ar(T-t) e^{-\gamma(\alpha^S(t)+\beta^S(t)W)} + K(W,t) - a\gamma(\dot{\alpha}^S(t)+\dot{\beta}^S(t) W) e^{-\gamma(\alpha^S(t)+\beta^S(t)W)} =
\]
\[
= -a\beta^S(t) e^{-\gamma(\alpha^S(t)+\beta^S(t)W)} + \frac{a}{2} ({\bf{\mu}}-\mu_0
\cdot{\bf 1})^\prime\Sigma^{-1}({\bf{\mu}}-\mu_0 \cdot{\bf 1}) e^{-\gamma(\alpha^S(t)+\beta^S(t)W)} +
\]
\begin{equation}\label{HJBexp}
 + \left(\mu_0 W-\alpha^S(t)-\beta^S(t) W + \frac{\hbox{ln}\, (a\gamma\beta^S(t))}{\gamma}\right) a\gamma\beta^S(t) e^{-\gamma(\alpha^S(t)+\beta^S(t)W)}
\end{equation}
where, from (\ref{B3}) and (\ref{consumoestandardexp}),
\begin{equation}\label{Wexp1}
K = E\left[\int_t^{T} \theta(s-t)\left[ r(s-t) - r(T-t)\right]
\left(-a\beta^S(s) e^{-\gamma (\alpha^S(s)+\beta^S(s)W(s))}\right) ds\right]\; .
\end{equation}
From the previous results, let us assume that $\beta^S(t)=\beta(t)$ (see (\ref{betastandardexp})); i.e., we verify the
integro-differential equation (\ref{HJBCI}) for $V^S(W,t)= -a e^{-\gamma(\alpha^S(t)+\beta(t)W)}$.
Let us calculate the contribution of the integral term $K(W,t)$. Note that equation
(\ref{budget2}) can be written as
$d W(s)= \left[(\mu_0-\beta(s))W+B(s)\right] ds + C(s) d{\bf \bar{z}}$,
$W(t)=W_t$, where $B(s)= \frac{1}{\gamma\beta(s)}({\bf{\mu}}-\mu_0
\cdot{\bf 1})^\prime\Sigma^{-1}({\bf{\mu}}-\mu_0 \cdot{\bf 1})-\alpha^S(s)+
\frac{\hbox{ln}\,(a\gamma\beta(s))}{\gamma}$
and
$C(s)=\frac{1}{\gamma\beta(s)}
({\bf{\mu}}-\mu_0
\cdot{\bf 1})^\prime\Sigma^{-1}\bar{\Sigma}$.
The solution is
\[
W(s)=e^{\int_t^s(\mu_0-\beta(\tau))d\tau}\left[ W_t+\int_t^s B(\bar{s})
e^{-\int_t^{\bar s}(\mu_0-\beta(\tau))d\tau}\, d\bar{s} +\right.
\]
\[
\left. + \int_t^s
e^{-\int_t^{\bar s}(\mu_0-\beta(\tau))d\tau} C(\bar{s})\,
d{\bf \bar{z}}_{\bar s}\right] =
\]
\[
=\frac{1}{\beta(s)}\left[ \beta(t) W_t+\int_t^s \beta({\bar s})
B(\bar{s})\, d\bar{s} + \int_t^s
\beta({\bar s}) C(\bar{s})\, d{\bf \bar{z}}_{\bar s}\right]\; ,
\]
where we have used that $e^{\int_t^s(\mu_0-\beta(\tau))d\tau} = \frac{\beta(t)}{\beta(s)}$.
Then (\ref{Wexp1}) becomes
\[
K(W_t,t)=-a e^{-\gamma(\alpha^S(t)+\beta(t)W_t)}E\left[\int_t^T\theta(s-t)[r(s-t)-r(T-t)]
\beta(s)\cdot\right.
\]
\[
\left.\cdot e^{-\gamma\left[\alpha^S(s)-\alpha^S(t)+
\int_t^s \beta(\tau) B(\tau)\, d\tau +
\int_t^s
\beta(\tau) C(\tau)\, d{\bf \bar{z}}_{\tau}\right]}\, ds\right] = a A(\alpha^S,t) e^{-\gamma(\alpha^S(t)+\beta^S(t)W_t)}
\]
and (\ref{HJBexp}) can be rewritten as
\[
-r(T-t) + A(\alpha^S,t) - \gamma\dot{\alpha}^S(t) + \beta(t) - \frac{1}{2} ({\bf{\mu}}-\mu_0
\cdot{\bf 1})^\prime\Sigma^{-1}({\bf{\mu}}-\mu_0 \cdot{\bf 1}) + \gamma\alpha^S(t) \beta(t) -
\]
\[
- \beta(t) \hbox{ln}\, (a\gamma\beta(t)) - \gamma\left(\dot{\beta}(t) +\mu_0 \beta(t) - \beta^2(t)\right) W = 0
\]
Since (\ref{betastandardexp}) is the solution to the differential equation $\dot{\beta}(t) +\mu_0 \beta(t) - \beta^2(t)=0$ with the boundary condition $\beta(T)=1$, the assumption $\beta^S(t)=\beta(t)$ is consistent with the integro-differential equation (\ref{HJBexp}). Note that, in this case, the integral term $K(W,t)$ does not contribute to the calculation of $\beta^S(t)$. With respect to $\alpha^S(t)$, it is the solution of a very complicated
integro-differential equation.

\subsection{Comparison between the different solutions}

First of all, we summarize the results obtained for the log-utility in Table 1.

\begin{table}[ht]
\begin{center}
\begin{tabular}{ccc}
\hline
{\bf Consumption rule} & \vline & {\bf Portfolio rule} \\ \hline
$c^{P}(t)=\frac{\theta(t) W_t }{a\theta(T) +\int_t^T\theta(s)\, ds}$
& \vline & ${\bf w}^P=\Sigma^{-1}({\bf{\mu}}-\mu_0\cdot{\bf 1})$ \\ \hline
$c^N(t)=\frac{W_t}{a\theta(T-t) +\int_t^T\theta(s-t)\,ds}$ & \vline & ${\bf w}^N=\Sigma^{-1}({\bf{\mu}}-\mu_0\cdot{\bf 1})$ \\ \hline
$c^S(t)=\frac{W_t}{a\theta(T-t) +\int_t^T\theta(s-t)\,ds}$ & \vline & ${\bf w}^S=\Sigma^{-1}({\bf{\mu}}-\mu_0\cdot{\bf 1})$ \\
\hline
\end{tabular}
\end{center}
\caption{Logarithmic utility function.}
\end{table}

A relevant property of the logarithmic utility function is that the portfolio rule is independent of the discount factor, and it is the same for the pre-commitment, naive and sophisticated solutions. Concerning the consumption rule, it coincides for naive and sophisticated agents. This is a remarkable property, in the sense that naive and sophisticated behaviors are completely different in nature. However, this result is not surprising, since it coincides with that obtained by Pollak (1968) for the Strotz's model.

With respect to the pre-commitment solution, it will be different in general to the naive and sophisticated solutions, unless $r(t)$ is constant. Let us denote by $\lambda^P$, $\lambda^N$ and $\lambda^S$ the propensity to consume for the pre-commitment, naive and sophisticated solutions, respectively ($c(t)=\lambda(t) W_t$).
Since $r(s)$ is nonincreasing, then $\int_0^s r(\tau)\, d\tau\geq \int_t^{t+s} r(\tau)\, d\tau = \int_0^{t+s} r(\tau)\, d\tau - \int_0^{t} r(\tau)\, d\tau$. Therefore, $\theta(s)\theta(t)\leq\theta(s+t)$. In particular, $\theta(T-t)\theta(t)\leq\theta(T)$ and $\theta(s-t)\theta(t)\leq\theta(s)$, so $\lambda^P(t)\leq\lambda^N(t)=\lambda^S(t)$, i.e., naive (and sophisticated) agents overconsume compared with the pre-commitment solution.

Unlike the Ramsey model (see Barro (1999)),
for the logarithmic utility function
the naive (or sophisticated) solution will be in general non-observationally
equivalent to the standard solution with some constant discount rate $\rho$.
Note that, for such observational equivalence, it is necessary and sufficient that
$\alpha(t)=\alpha^S(t)$, for some $\rho$, i.e., from (\ref{consumostandard})
and (\ref{logsophisticated}), $\left( a-\frac{1}{\rho}\right) e^{-\rho(T-t)}
+ \frac{1}{\rho} = a\theta(T-t) +\int_t^T\theta(s-t)\,ds$. If $a=0$, then by differentiating
with respect to $t$ we obtain that $\theta(T-t)=e^{-\rho(T-t)}$
and the discount factor is constant. For $a\neq 0$, if $\theta(T-t)\neq e^{-\rho(T-t)}$, by
solving $a$ we obtain
\begin{equation}\label{nonequiv}
a=\frac{\frac{1}{\rho}\left(1-e^{-\rho(T-t)}\right)
- \int_t^T\theta(s-t)\,ds}{\theta(T-t)-e^{-\rho(T-t)}}\; .
\end{equation}
By defining $x(t)=\frac{1}{\rho}\left(1-e^{-\rho(T-t)}\right)
- \int_t^T\theta(s-t)\,ds$, the above equation becomes
$a=\frac{x(t)}{\dot{x}(t)}$, therefore $x(t)=A e^{t/a}$. By identifying $\dot{x}(t)$
with the denominator in the right hand term in (\ref{nonequiv})
we obtain $\theta(T-t)=\frac{A}{a}e^{t/a}+e^{-\rho(T-t)}$.
Hence, there will be observational equivalence
with a standard model with constant discount rate $\rho$ if,
and only if, $a\neq 0$ and $\theta(t)$ is a linear combination of two exponentials
with constant discount rates $\rho$ and $1/a$.

We recover observational equivalence in the limit $T\to\infty$. In this case, the propensity to consume in the case of a constant discount rate $\rho$ is $\lambda=\rho$. Under non-constant discounting, if we assume that $\lim_{t\to\infty}r(t)=\bar{r}>0$, the propensity to consume of naive and sophisticated agents becomes constant, $\lambda^N=\lambda^S=1/\int_0^\infty\theta(\tau)d\tau$. Then, in the limit $T\to\infty$, we obtain observational equivalence with effective discount rate $\rho=[\int_0^\infty\theta(\tau)d\tau]^{-1}$.

Table 2 presents the results obtained for a potential utility function.

\begin{table}[ht]
\begin{center}
\begin{tabular}{ccc}
\hline
{\bf Consumption rule} & \vline & {\bf Portfolio rule} \\ \hline
$c^P(t)=\frac{e^{-\delta^p(T-t)} \theta(t) W}{\theta(T)\left( a+
\int_t^T \left(\frac{\theta(s)}{\theta(T)}e^{-\delta^p(T-s)}\right)^{\frac{1}{1-\gamma}}
\,
ds\right)}$ & \vline & ${\bf w}^P=\frac{1}{1-\gamma}\Sigma^{-1}({\bf{\mu}}-\mu_0\cdot{\bf 1})$ \\ \hline
$c^N(t)=\frac{e^{-\delta^p(T-t)} W}{\theta(T-t)\left(a+
\int_t^T \left(\frac{\theta(s-t)}{\theta(T-t)}e^{-\delta^p(T-s)}\right)^{\frac{1}{1-\gamma}}
\,
ds\right)}$ & \vline & ${\bf w}^N=\frac{1}{1-\gamma}\Sigma^{-1}({\bf{\mu}}-\mu_0\cdot{\bf 1})$ \\ \hline
$c^S(t)=\left(\alpha^S\right)^{-\frac{1}{1-\gamma}} W_t$, $\alpha^S(t)$ given by (\ref{sofpot}) & \vline & ${\bf w}^S=\frac{1}{1-\gamma}\Sigma^{-1}({\bf{\mu}}-\mu_0\cdot{\bf 1})$ \\
\hline
\end{tabular}
\end{center}
\caption{Potential utility function.}
\end{table}

Again, the portfolio rule is independent of the discount factor, and
coincides for the three solutions. However,
the coincidence of the consumption rule of naive and sophisticated agents
in the logarithmic case is not preserved. A possible interpretation for this result can be found in
Mar\'{\i}n-Solano and Navas (2008) (see also Barro (1999)). Since the integral in equation (\ref{sofpot}) disappears in the logarithmic case (which is the limit when $\gamma\to 0$ of the utility function $u(c)=(c^\gamma-1)/\gamma$), the sophisticated $t$-agent decides his optimal consumption at time $t$ without being affected by his future selves and,
in some way, behaving as an homogeneous decision-maker, similarly to a naive agent.

In the limit $T\to \infty$, if and $r(t)>\delta^p$, the propensities to consume $\lambda^P$, $\lambda^N$ are
\begin{equation}\label{propensityP}
\lambda^P=\frac{\tilde{\theta}(t)}{\int_t^\infty\left[\tilde{\theta}(s)\right]^{\frac{1}{1-\gamma}}\, ds}\; ,
\end{equation}
where $\tilde{\theta}(s)=e^{-\int_0^s\tilde{r}(\tau)d\tau}$, with $\tilde{r}(\tau)=r(\tau)-\delta^p$, and
\begin{equation}\label{propensityN}
\lambda^N=\frac{1}{\int_0^\infty\tilde{\theta}(\tau)^{\frac{1}{1-\gamma}}\, d\tau}
= \frac{\bar{r}-\delta^p}{1-\gamma-\int_0^\infty\left[ r(\tau)-\bar{r}\right]
\left[\tilde{\theta}(\tau)\right]^{\frac{1}{1-\gamma}}\, d\tau}\; .
\end{equation}
For $t=0$, the propensities to consume given by the naive and the precommitment solutions coincide, but $(\lambda^P)^\prime(0)\leq 0$ ($(\lambda^N)^\prime(0)=0$), as expected. In general, since $\int_0^\infty\theta(\tau)^{\frac{1}{1-\gamma}}\, d\tau=[
1-\gamma+\int_0^\infty( r(0)-r(\tau))(\tilde{\theta}(\tau))^{\frac{1}{1-\gamma}}\, d\tau]/(r(0)-\delta^p)$, from (\ref{propensityN}) we get $(\bar{r}-\delta^p)/(1-\gamma)\leq\lambda^N\leq (r(0)-\delta^p)/(1-\gamma)$. With respect to $\lambda^S$,
\begin{equation}\label{propensityS}
\lambda^S=\frac{\bar{r}-\delta^p}{1-\gamma-\int_0^\infty\left[ r(\tau)-\bar{r}\right]
\theta(\tau) e^{\gamma\Delta\tau}\, d\tau}
\end{equation}
and therefore $\lambda^S\geq(\bar{r}-\delta^p)/(1-\gamma)$. Moreover, since the propensity to consume under an instantaneous discount rate $\rho$ is $\lambda=(\rho-\delta^p)/(1-\gamma)$, we obtain observational equivalence
for naive and sophisticated agents (with different effective discount rates
$\rho=(1-\gamma)\lambda^N+\delta^p$ and $\rho=(1-\gamma)\lambda^S+\delta^p$, respectively).

Finally, note that the coincidence of the portfolio rule for the different behaviors of the decision maker in the case of CRRA utility functions is no longer satisfied for more general HARA utility functions, such
as the exponential (CARA) utility function. Although in the exponential case
the expression of the portfolio rule (as a function of $(W,t)$) is the same for an agent with constant discount rate
and the different solutions with non-constant discounting (they are all given by
(\ref{portfoliostandardexp}), with $\beta(t)=\beta^P(t)= \beta^N(t)=\beta^S(t)$),
this property does not imply that the portfolio rule is independent of the discount
factor, since the evolution of $W(t)$ depends on the values of $\alpha(t)$,
$\alpha^P(t)$, $\alpha^N(t)$ and $\alpha^S(t)$, respectively, and all these
functions do not coincide, in general. The consumption rule is also different.
It is easy to check that, at time $t=0$, $\alpha^P(0)=\alpha^N(0)$ but $(\alpha^P)^\prime(0)\leq(\alpha^N)^\prime(0)$, hence $(c^N)^\prime(0)\geq (c^P)^\prime(0)$.

\section{Concluding remarks}

In this paper we study the problem of searching for
optimal/equilibrium rules in the case where decision-makers have
time-inconsistent preferences, within a stochastic framework. For
the so-called sophisticated agents, we derive
a modified HJB equation which extends the equation
for a deterministic problem (see Karp (2007) and Mar\'{\i}n-Solano and
Navas (2008)). This equation is obtained, first, heuristically, and it is
mathematically justified later on, by following a different approach.
Although this modified HJB equation seems to be too complicated
in general, we illustrate with several examples how it can be managed
in order to obtain information about the solution. In particular, this
modified HJB equation is used in order
to solve (for some utility functions) the classical consumption
and portfolio rules model when the instantaneous discount
rate of time preference is non-constant.

A relevant result is that
for the CRRA (logarithmic and potential) utility functions, the
portfolio rule coincides for the pre-commitment, naive and sophisticated
solutions. Moreover, it is independent of the discount factor, and thus
coincides with the standard solution when the discount factor is an
exponential with constant discount rate. This property is no longer
satisfied for more general utility functions, such as the CARA
(exponential) function.

With respect to the consumption rule, it is proved that, in the
log-utility case, it coincides for naive and sophisticated agents.
This is a remarkable property which extends to the Merton's model
a similar result already announced in Pollak (1968) for the
Strotz model. This coincidence is no longer satisfied for
more general utility functions. In the log-utility case, the
observational equivalence problem first studied for the Ramsey model
in Barro (1999) is analyzed, with a negative
answer except for a very particular form of the discount factor.

Possible extensions of the results in the paper include several
financial and actuarial applications, such as contribution and
portfolio selection in pension funding (see, e.g., Josa-Fombellida
and Rinc\'{o}n-Zapatero (2008) and references therein).


\begin{thebibliography}{}


\bibitem{Ai92} Ainslie, G.W. (1992). Picoeconomics. Cambridge University
Press, Cambridge, UK.

\bibitem{Ba99} Barro, R.J. (1999). Ramsey meets Laibson in the neoclassical
growth model. Quarterly Journal of Economics 114, 1125-1152.

\bibitem{Bo90} Boyd, J.H. III (1990). ``Symmetries, Dynamic Equilibria,
and the Value Function'', in Conservation Laws and Symmetry: Applications
to Economics and Finance, R. Sato and R.V. Ramachandran (eds.),
Kluwer Academic Publishers, Boston.

\bibitem{Ch04} Chang, F.-R. (2004). Stochastic Optimization in
Continuous Time. Cambridge University
Press, Cambridge, UK.

\bibitem{EL08} Ekeland, I. and Lazrak, A. (2008). Equilibrium policies
when preferences are time inconsistent. arXiv:0808.3790v1 [math.OC].

\bibitem{EL08b} Ekeland, I. and Pirvu, T.A. (2008a). Investment and Consumption
without Commitment. arXiv:0708.0588v1 [q-fin.PM].

\bibitem{EL08c} Ekeland, I. and Pirvu, T.A. (2008b). On a Non-Standard Stochastic Control Problem.
arXiv:0806.4026v1 [q-fin.PM].

\bibitem{FR75} Fleming, W.H. and Rishel, R.W. (1975). Deterministic and Stochastic Optimal Control. Springer, New York, USA.

\bibitem{FS06} Fleming, W.H. and Soner, H.M. (2006). Controlled Markov Processes
and Viscosity Solutions. Springer, New York, USA.

\bibitem{Gr07} Grenadier, S.R. and Wang, N. (2007). Investment under
uncertainty and time-inconsistent preferences. Journal of Financial
Economics 84, 2-39.

\bibitem{HL05} Harris, C., Laibson, D. (2008). Instantaneous
Gratification. Working Paper.

\bibitem{Jo08} Josa-Fombellida, R. and Rinc\'{o}n-Zapatero, J.P. (2008).
Mean-variance portfolio and contribution selection
in stochastic pension funding.
European Journal of Operational Research 187, 120-137.

\bibitem{Ka07} Karp, L. (2007). Non-constant discounting in continuous time.
Journal of Economic Theory 132, 557-568.

\bibitem{KD01} Kushner, H.J. and Dupuis, P. (2001). Numerical Methods for
Stochastic Control Problems in Continuous Time. Springer, New York, USA.

\bibitem{La97} Laibson, D. (1997). Golden Eggs and Hyperbolic Discounting.
Quarterly Journal of Economics 112, 443-477.

\bibitem{Lo92} Loewenstein, G. and Prelec, D. (1992). Anomalies in
intertemporal choice: evidence and an interpretation. Quarterly Journal
of Economics 57, 573-598.

\bibitem{Ma08} Mar\'{\i}n-Solano, J. and Navas, J. (2009). Non-constant
discounting in finite horizon: The free terminal time case. Journal of Economic
Dynamics and Control 33 (3), 666-675.

\bibitem{Me69} Merton, R.C. (1969). Lifetime portfolio selection under
uncertainty: the continuous time case. Review of Economics and Statistics
51, 247-257.

\bibitem{Me71} Merton, R.C. (1971). Optimum consumption and portfolio
rules in a continuous time model. Journal of Economic Theory 3, 373-413.

\bibitem{Ph68} Phelps, E.S., Pollak, R.A. (1968). On Second-best
National Saving and Game-Equilibrium Growth. Review of Economic Studies
35, 185-199.

\bibitem{Po68} Pollak, R.A. (1968). Consistent Planning. Review of Economic
Studies 35, 201-208.

\bibitem{Ra28}  Ramsey, F. (1928). A Mathematical Theory of Saving.
Economic Journal 38, 543-559.

\bibitem{Sr56} Strotz, R.H. (1956). Myopia and Inconsistency in Dynamic
Utility Maximization. Review of Economic Studies 23, 165-180.

\bibitem{Th81} Thaler, R. (1981). Some empirical evidence on dynamic
inconsistency. Economics Letters 8, 201-207.

\bibitem{To08} Tomak, K. and Keskin, T. (2008). Exploring the trade-off between immediate gratification
and delayed network externalities in the consumption of information goods.
European Journal of Operational Research 187, 887-902.


\end{thebibliography}
\end{document}